\documentclass[conference]{IEEEtran}
\usepackage{cite}
\usepackage{hyperref}
\usepackage{amsmath,amssymb,amsfonts}
\usepackage{algorithmic}
\usepackage{graphicx}
\usepackage{textcomp}
\usepackage{xcolor}
\usepackage{subcaption}

\makeatletter
\newcommand{\linebreakand}{%
  \end{@IEEEauthorhalign}%
  \vspace{1em}\par
  \begin{@IEEEauthorhalign}%
}
\makeatother

\def\BibTeX{{\rm B\kern-.05em{\sc i\kern-.025em b}\kern-.08em
    T\kern-.1667em\lower.7ex\hbox{E}\kern-.125emX}}
\begin{document}

\title{Accelerating Quantum Tensor Network Simulations with Unified Path Variations and Non-Degenerate Batched Sampling\\
}

\author{\IEEEauthorblockN{1\textsuperscript{st} Taylor Lee Patti}
\IEEEauthorblockA{\textit{NVIDIA} \\
Santa Clara, USA \\
tpatti@nvidia.com}
\and
\IEEEauthorblockN{2\textsuperscript{nd} Paavai Pari}
\IEEEauthorblockA{\textit{NVIDIA} \\
Santa Clara, USA  \\
ppari@nvidia.com}
\and
\IEEEauthorblockN{3\textsuperscript{rd} Yang Gao}
\IEEEauthorblockA{\textit{NVIDIA} \\
Santa Clara, USA \\
yangg@nvidia.com}
\and
\IEEEauthorblockN{4\textsuperscript{th} Azzam Haidar}
\IEEEauthorblockA{\textit{NVIDIA} \\
Santa Clara, USA \\
ahaidarahmad@nvidia.com}
\linebreakand
\and
\IEEEauthorblockN{5\textsuperscript{th} Thien Nguyen}
\IEEEauthorblockA{\textit{NVIDIA} \\
Santa Clara, USA \\
thiennguyen@nvidia.com}
\and
\IEEEauthorblockN{6\textsuperscript{th} Tom Lubowe}
\IEEEauthorblockA{\textit{NVIDIA} \\
Santa Clara, USA \\
tlubowe@nvidia.com}
\and
\IEEEauthorblockN{7\textsuperscript{th} Daniel Lowell}
\IEEEauthorblockA{\textit{NVIDIA} \\
Santa Clara, USA \\
dlowell@nvidia.com}
\and
\IEEEauthorblockN{8\textsuperscript{th} Brucek Khailany}
\IEEEauthorblockA{\textit{NVIDIA} \\
Santa Clara, USA \\
bkhailany@nvidia.com}}

\maketitle

\begin{abstract}
Quantum trajectory methods reduce the computational overhead of simulating noisy quantum systems, approximating them with $m$ stochastically sampled $2^n$-entry quantum statevectors rather than exact $2^{2n}$-entry density matrices. Recently, Pre-Trajectory Sampling with Batched Execution (PTSBE) has dramatically increased the data collection rate of these methods. While statevector PTSBE has demonstrated data collection speedups of over $10^6 \times$, tensor network implementations only achieved $\sim 15 \times$ speedup. This comparatively modest tensor network advantage stemmed from 1) contraction path recalculations,  2) sequential tensor network sampling, and 3) inflexible/unoptimized contraction hyperparameters. In this manuscript, we increase PTSBE's tensor network data collection rate to more than $10^8\times$ that of traditional trajectories methods by developing 1) error-independent unified path variation, 2) non-degenerate tensor network sampling, and 3) a flexible/optimized contraction framework. While our methods are particularly powerful for accelerating non-proportional sampling, we also demonstrate a more than $1000\times$ speedup for more general quantum simulations.
\end{abstract}


\section{Introduction}
Quantum computing is the focus of growing scientific excitement and investment \cite{ruane2025quantum,cassemiro2025}. While such devices appear promising for applications in chemistry \cite{mcardle2020}, material science \cite{alexeev2024quantumcentric,akanbi2025minireview}, and even traditional computation \cite{dalzell2025quantum,devadas2025quantum}, the path towards quantum computers is complicated by a foundational dilemma: the development of quantum computers is desirable because of their high classical complexity, but this computational complexity makes quantum computers exquisitely difficult to simulate. The ramifications of this dilemma are even more stark in the age of AI, as many of the AI techniques that stand to advance quantum information not only have high data requirements \cite{maslej2025aiindex}, they also benefit greatly from high data quality \cite{albalak2024survey}. Nevertheless, the integration of AI into the quantum sciences is considered crucial \cite{alexeev2025artificial}, particularly in fields such as quantum device design and quantum error correction \cite{wang2024artificial}.

Quantitatively speaking, the \textit{general} complexity of exactly simulating $n$ ideal quantum bits (qubits) scales as a $2^n$-entry vector, often referred to as a \textit{statevector}. Moreover, the complexity of simulating $n$ noisy (realistic) qubits is quadratically more complex, scaling as a $2^{2n}$-entry matrix \cite{scully1997quantum,campaioli2024quantum}. Various alternative methods of quantum simulation that do not inherently scale exponentially with $n$ exist, such as Clifford \cite{gidney2021stim}, Near-Clifford \cite{bravyi2016improved,bravyi2019simulation}, PauliProp \cite{rudolph2025pauli}, and others (see Sec. \ref{subsec:noisy_simulation} for a more complete discussion), but tensor networks are arguably the oldest and most general alternative to universal statevector simulations \cite{berezutskii2025tensor}. In particular, they can carry out efficient, exact simulations in systems where there are relatively many qubits and few quantum gates (quantum logical operations), as unlike statevectors, their complexity grows only polynomially in the former, albeit exponentially in the latter \cite{markov2008simulating}. However, the complexity of tensor network simulations also greatly increases when moving from the coherent to the noisy regime.

\begin{figure*}[]
    \centering
    \includegraphics[width=1.025\textwidth, page=1  ]{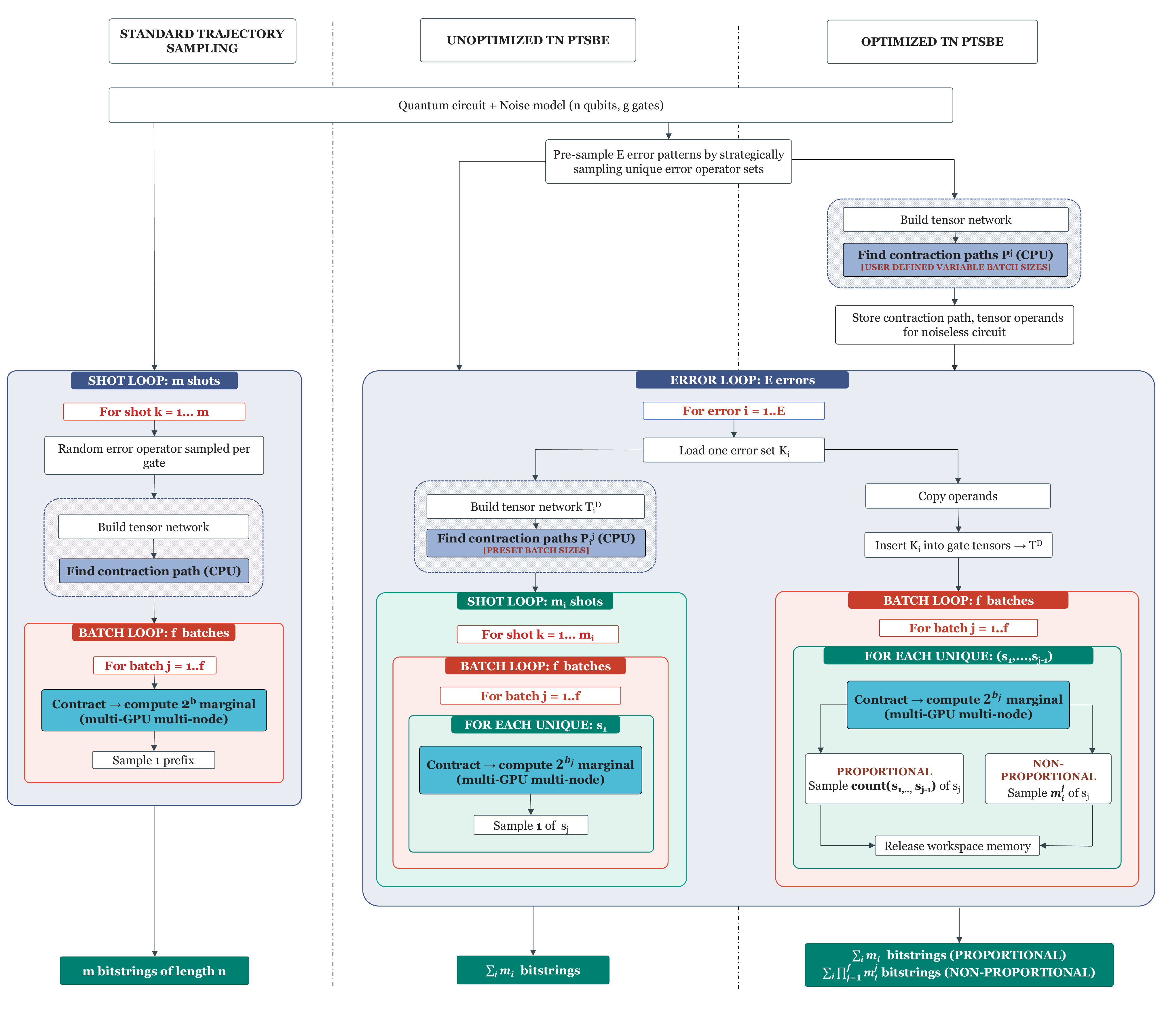}
    \caption{Diagram of trajectory sampling techniques for tensor networks. (Left) Traditional trajectory simulations must recompute the trajectory paths $P_i^j$ for each error set $K_i$ with fixed batch sizes $b$, and they harvest only one shot (quantum measurement) per full series of quantum contractions $j$ \cite{cuda-q}. (Center) Unoptimized PTSBE makes some improvement on this scheme by finding the contraction paths $P_i^j$ for each $K_i$ just one time, however it still must find $E$ distinct sets of such paths, it retains fixed batch size $b$, and it is limited to one shot per full contraction loop \cite{patti2025augmenting_sc}. In contrast, this manuscript's optimized version of PTSBE 1) uses a single, stored contraction path through a lightweight tensor merge operation, 2) carries out zero degenerate contractions by batch-processing intermediate bitstrings $(s_1, \dots, s_{j-1})$ and (in the non-proportional case) capturing massive amounts of quantum shots from the final batch $B_f$, and 3) develops a flexible interface for optimizing over contraction batch sizes $b_j$. We highlight that optimized tensor network PTSBE, like the other trajectory simulation methods discussed, is embarrassingly parallelizable up to $E$ distinct GPUs.}
    \label{fig:algorithm_diagram}
\end{figure*}

To address this overwhelming overhead for noisy systems, so-called ``quantum trajectory methods'' have been devised \cite{dalibard1992wave,dum1992monte,daley2014,isakov2021simulations}. These forgo the construction of a full $2^{2n}$-entry density matrix, substituting instead an ensemble of $m$ $2^n$-entry statevectors. While still computationally intensive, this approximation is highly favorable when $m \ll 2^n$, a condition that is easily satisfied. Recently, the data collection efficiency of trajectory methods for statevector simulation was increased by up to six orders of magnitude by the development of Pre-Trajectory Sampling with Batched Execution (PTSBE) \cite{patti2025augmenting_sc}, however the tensor network version of PTSBE demonstrated just a $15\times$ speedup due to a lack of universal path finding, conditional sampling, and simulator flexibility/optimization. Moreover, the resolution of these three computational bottlenecks stands to accelerate quantum tensor network simulations more broadly, as it would remove costly and redundant work from a much wider array of iterative and many-shot quantum sampling workloads.

In this manuscript, \textbf{\textit{we introduce three unique innovations}} for dramatically accelerating quantum tensor network trajectory simulations:

\begin{enumerate}
    \item \textbf{Unified Path Variations (UPV) -} a framework for finding contraction paths that is universal for entire families of quantum tensor networks, such that these expensive subroutines can be done once instead of thousands to millions of times. The repeated contraction path searches that UPV eliminates are ubiquitous in, but not limited to, noisy quantum system simulation.
    \item \textbf{Non-Degenerate Batched Sampling (NBS) -} a series of methods for removing repeated (degenerate) operations from tensor network partial sampling that also allows users to collect broad swaths of shot (measurement) data from the simulation.
    \item \textbf{Interface flexibility and optimization -} the design of a more flexible interface for tensor network simulations of noisy quantum systems, which we in turn use to find markedly more optimal sampling hyperparameters.
\end{enumerate}

In what follows, we implement these contributions in an end-to-end simulation pipeline using the cuQuantum cuTensorNet library \cite{bayraktar2023cuquantumsdkhighperformancelibrary}, CuPy \cite{nishino2017cupy}, and a Qiskit intermediate representation \cite{qiskit2024} (the intermediate representation was only used to generate uncontracted tensor networks from quantum circuit specifications, it was not used for any sizable computational task). We go on to demonstrate dramatic speedups compared to the traditional tensor network trajectory simulator in CUDA-Q \cite{cuda-q}, namely a $10^8 \times$ data collection speedup for non-proportional PTSBE and $10^3 \times$ data collection speedup for proportional PTSBE. We further highlight that proportional PTSBE adheres to traditional quantum statistics, meaning that our implementation is a general acceleration of tensor network trajectory simulations, without added algorithmic approximation or restrictions.

\section{Background}
\label{sec:background}

\begin{figure*}[]
    \centering
    \includegraphics[width=1.05\textwidth, page=1]{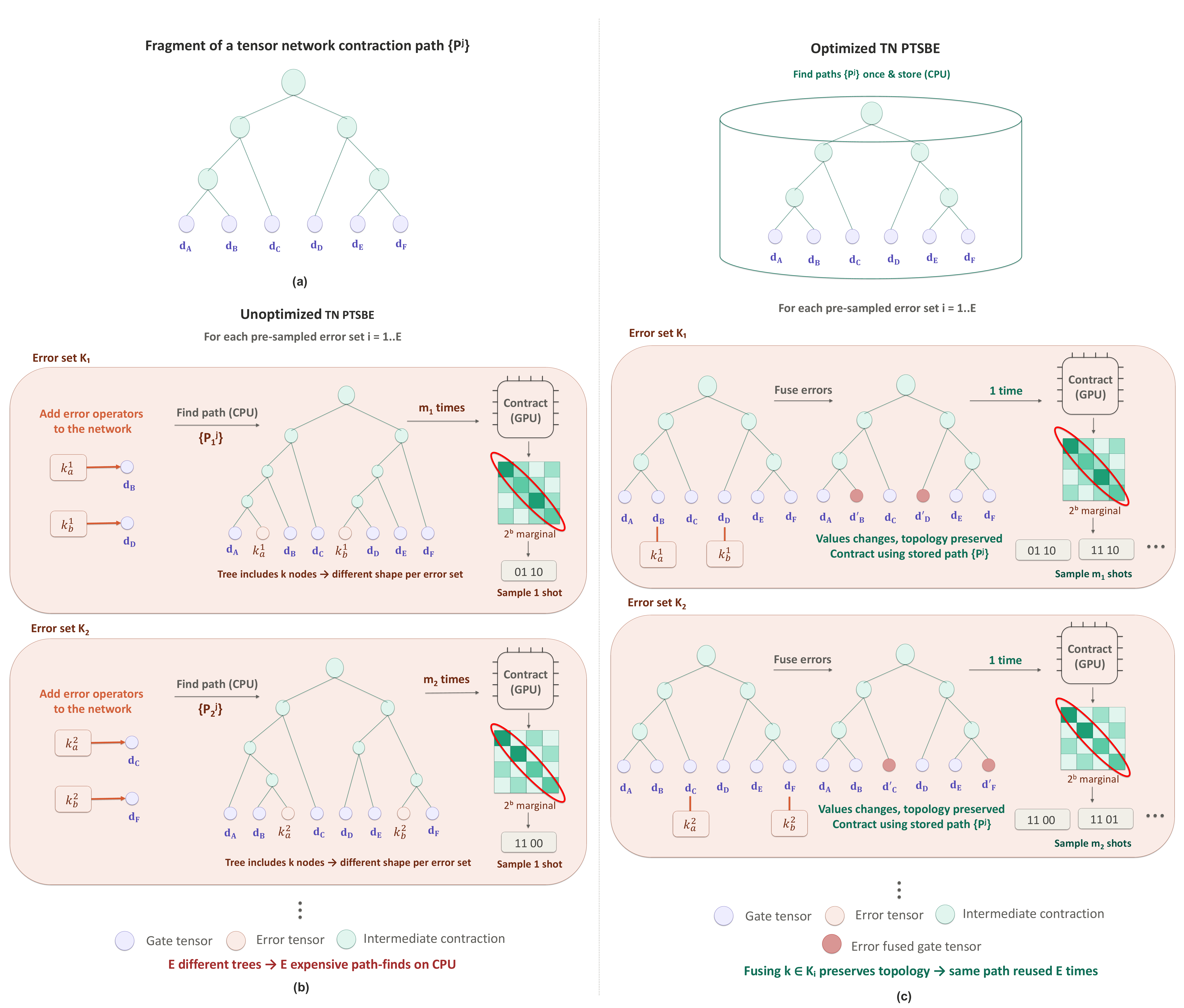}
    \caption{a) Fragment of a quantum circuit contraction path. b) Repeated contraction path finding for unoptimized TN PTSBE. With the traditional error-insertion method, for each error set $K_i$, a distinct tensor network topology is generated and a new, suitable contraction path is found. The process for typical trajectory simulations is similar, but even more computationally intensive, as it requires the computation of a new contraction path for every \textit{shot}, not just every error set. c) In contrast, optimized TN PTSBE enjoys UPV, a method for reusing a single contraction path for all error combinations by calculating and storing the error-free contraction paths $P^j$, and then reusing this path by integrating error operands $k^i_p$ into the corresponding coherent operands $d_l$, a lightweight operation that preserves tensor network structure.}
\label{fig:contraction_path_reuse}
\end{figure*}

In order to contextualize the extensive contributions of this manuscript, we first detail previous contributions in the field.

\subsection{Noisy Quantum Simulation Techniques and Applications}
\label{subsec:noisy_simulation}

Simulators capable of modeling noisy quantum systems are dearly needed for the development of quantum computers. The realistic noise sources for these devices are diverse, including environmental coupling \cite{breuer2002theory}, gate errors \cite{bharti2022nisqAlgorithms,bartolomeo2023noisygates}, and material defects \cite{martinis2005decoherence}. While modeling and understanding these errors is important to virtually every aspect of quantum computer design, noisy simulation for quantum error correction \cite{terhal2015quantum,roffe2019quantum}, a branch of quantum study that seeks to achieve fault-tolerant computation \cite{steane2003threshold,fowler2009high_threshold}, is of particular interest. Moreover, many AI-based quantum error-correcting workloads require large amounts of \textit{shot} (quantum measurement) data to train, such as AI decoders \cite{bausch2024learning,torlai2017neural}.

However, as discussed in the introduction, the simulation of noisy quantum systems is even more challenging to simulate than coherent quantum systems, which are already of exponential complexity. Specifically, general simulations of noisy quantum systems require a $2^{2n}$-entry matrix rather than just a $2^n$-entry statevector \cite{scully1997quantum,campaioli2024quantum}. Various approximations and algorithmic tradeoffs have been developed to address this enormous overhead, including Clifford gate-like simulations \cite{gottesman1998heisenberg,bravyi2016improved,bravyi2019simulation,gidney2021stim,rudolph2025pauli}, approximate tensor networks \cite{white1992density,baiardi2019large}, and system reduction techniques \cite{schrieffer1966relation,brion2007adiabatic,yadav2023legate,chakraborty2025gpu}. Various graph-based and approximate tensor simulators have also been developed \cite{dd_densitymat, intel_qs, isakov2021simulations, dmsim, mpdo_sim, dd_noisy_sim, aer_trajectory}.

While each of these methods offers its own benefits and drawbacks, quantum trajectory methods are of particular utility for HPC simulations of large quantum systems.

\subsection{Quantum Trajectories Methods}
Quantum trajectories methods are ideal for large, high-fidelity, exact-gate noisy quantum simulations of arbitrary $T$-gate number \cite{dalibard1992wave,dum1992monte,daley2014}. They are also ideal for HPC simulations, as they can be scaled to fit nearly arbitrarily large hardware due to their embarrassing parallelizability over distinct trajectories and the substantial software infrastructure for multi-GPU statevector and state-like quantum simulation. As a result, quantum trajectory simulations have been widely used and implemented in various software packages \cite{isakov2021simulations,cuda-q}. A schematic of these standard trajectory sampling methods for tensor networks is given in Fig. \ref{fig:algorithm_diagram} (left). In order to collect $m$ quantum shots (quantum measurements) from an $n$-qubit system, trajectory methods repeat an expensive tensor network sampling loop $m$ times. For each of the $m$ iterations, all potential error operator $k$ are sampled in accordance with their respective probability, resulting in some subset of quantum errors $K_i = \{k_p^i\}$ being selected. An uncontracted tensor network $T^D_i$, containing both the deterministic coherent gates $D = \{d_l\}$ and this stochastically sampled error subset $K_i$, is then generated. Gathering quantum shot data for $T^D_i$ must be accomplished via a series of CPU-based optimizations to find contraction paths $P_i^j$, where the index $j$ enumerates the division of the $n$ qubits into batches of $b$ qubits. While this batching is common for sampling from quantum tensor networks, trajectory method APIs typically only support default qubit batches $B_j$ of size $b_j$, where $b_j$ is not necessarily optimal. Generally, $b_j = n//b$ for non-final batches ($j<f$) and $b_f = n\%b$ for the final batch $B_f$, where $f=\text{ceil}(n/b)$, $//$ denotes integer division, and $\%$ denotes division remainder. For instance, CUDA-Q sets $b=24$ \cite{cuda-q}.

Shots (quantum measurements) are sampled iteratively through batches $B_j$ (Fig. \ref{fig:algorithm_diagram}, left), such that the outcomes of all preceding shot bitstrings $s_1,..., s_{j-1}$ are inserted into the network to partially project the state. This piece-wise conditional sampling is needed for memory management, as each distribution that is extracted for sampling is of dimension $2^b$. This process leads to a concatenated bitstring set $(s_1,...,s_f)$. We emphasize the completely iterative nature of vanilla PTSBE. That is, as all contraction paths are calculated de novo and all shots are taken individually, all computational steps must be repeated for the collection of each random shot. 

\subsection{Unoptimized Tensor Network PTSBE}
\label{subsec:unoptimized}

Recently, PTSBE methods have been introduced to remove degenerate work from statevector and tensor network trajectory simulations \cite{patti2025augmenting_sc}. In addition to marked speedup for exact trajectory sampling, PTSBE can achieve orders of magnitude data collection speedup for sampling paradigms that do not require strict adherence to quantum sampling statistics, such as data collection for ML applications in noisy quantum systems \cite{alexeev2024quantumcentric}. While statevector data collection speedups reached up to $10^6 \times$, tensor network implementations enjoyed a much lower advantage of $\sim 15\times$. This comparatively humble speedup stems from various implementation limitations, as detailed below.

The center panel of Fig. \ref{fig:algorithm_diagram} illustrates this original \textit{unoptimized} tensor network PTSBE implementation, its benefits, and its limitations. Rather than sampling one error set per shot at runtime, $E$ such error sets $K_i$ are pre-sampled and stored prior to the construction or evaluation of any quantum tensor network $T^D_i$. This pre-trajectory sampling (PTS) is done according to a user-specified rule (e.g., proportional with error probabilities) and the corresponding number of shots $m_i$ are likewise assigned in accordance with the specified rule. Such sampling is of low-degree polynomial-complexity, and PTSBE is devised to factorize this lightweight work from the traditional trajectories simulation protocol with virtually no increase in computational overhead. As further illustrated in Fig. \ref{fig:contraction_path_reuse}, the corresponding tensor network $T^D_i$ is then constructed with the coherent (deterministic) gates $D$ and the noise gates $\{k_p^i\}$ for each $K_i$, and then the $f$ contraction paths $P_i^j$ on fixed qubit batch sizes $b$ are calculated and stored. Finally, $m_i$ shots are calculated by carrying out all $f$ contraction paths $m_i$ times, with each set of contractions obtaining just one shot at a time. We highlight that the advantage of this original tensor network PTSBE workflow was the reutilization of contraction paths $P_i^j$, however this original method remains limited due to 1) the recalculation of $P_i^j$ for each error set $K_i$, 2) the fixed nature of batch size $b$, and 3) the fully sequential shot collection process. More details are available in \cite{patti2025augmenting_sc}.

\section{Methods - Optimized Tensor Network PTSBE}\label{sec:methods}

\begin{figure}[t]
\centerline{\includegraphics[width=\columnwidth]{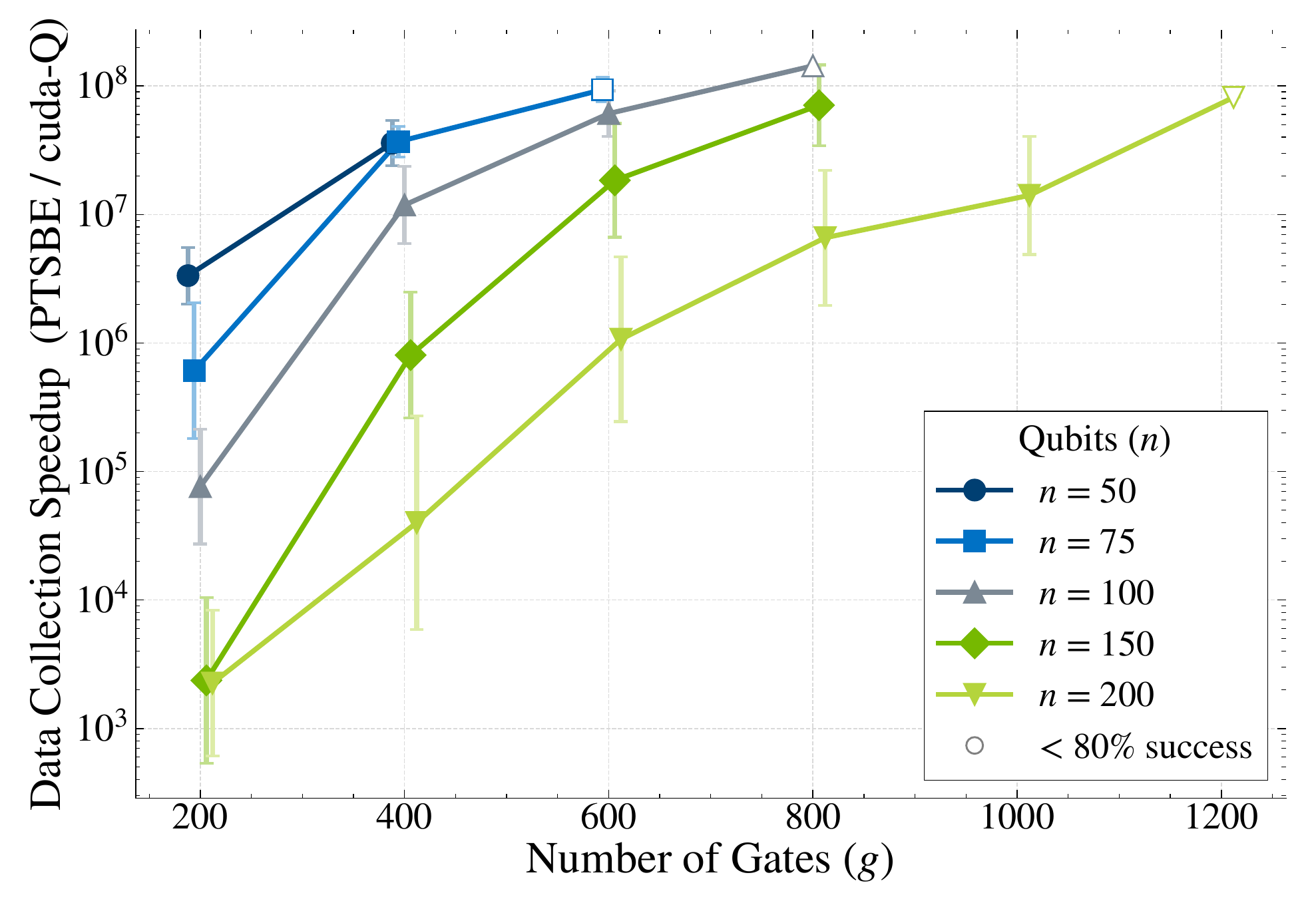}}
\caption{Data collection speedup for optimized non-proportional tensor network PTSBE vs traditional CUDA-Q trajectories simulation. PTSBE speedup is marked in all regimes, reaching up to more than $10^8 \times$ speedup. Speedup grows with circuit depth (relative to qubit number) as deeper circuits generally have more states populated and thus benefit more from batched shot harvesting.}
\label{fig:speedup_vs_gates}
\end{figure}

We now present this manuscript's contribution, an optimization of tensor network PTSBE that removes the two largest sources of degenerate work: 1) repeated tensor network path calculations and 2) single-shot tensor network calculations. In addition to these two algorithmic improvements, we also devise an interface that exposes bespoke qubit batch-sizes for trajectory sampling, the flexibility of which allows optimization and additional speedup. We further highlight that optimized tensor network PTSBE is an embarrassingly parallelizable HPC algorithm, as it can, with only the minimal overhead of distributing small uncontracted tensors and a set of lightweight contractions paths, scale to as many GPUs as error sets $E$ studied.

\subsection{Unified Path Variations}
\label{subsec:upv}

As discussed in Sec. \ref{subsec:unoptimized}, previous tensor network implementations of PTSBE were able to reduce the overhead of contraction path by finding and caching $P_i^j$, then reusing it for all $m_i$ pre-allocated shots, rather than recalculating it every time, as is done in traditional trajectory methods. For each $K_i$, unoptimized PTSBE formed a distinct tensor network by inserting the distinct error operators $\{k_p^i\}$ alongside the deterministic gates $\{d_l\}$. As these contraction paths are generally unique, each such path must be found independently, an expensive CPU-based calculation.

We eliminate this repetitive computational overhead, by introducing unified path variations (UPV), as illustrated in Fig. \ref{fig:contraction_path_reuse} (right). In UPV, we make the very common quantum physical modeling assumption that errors are preceded or succeeded by a gate of the same size (e.g., single-qubit, two-qubit) and location. Using this assumption, we see that we can insert each $K_i$ into a copy of the tensors $D$ and then tensor merge (contract) them into the nearest tensor $d_l$. As long as all of the tensors are modeled as full-rank (or dummy-ranks are inserted so it can be modeled as such), this lightweight tensor merge operation produces a tensor network with the same tensor network structure (the same operand number, shape, topology, and rank) as $D$ itself, allowing us to now compute a highly optimized and general use contraction path set $P^j$ for use for \textit{all errors} and \textit{all shots}, rendering contraction path search time virtually negligible when compared to the many error sets $K_i$ and shots $m_i$ taken in even a modestly-sized trajectory methods sampling study.

\subsection{Non-Degenerate Batched Sampling}
\label{subsec:nbs}

In previous trajectory sampling implementations, including both standard and unoptimized PTSBE, a single shot was garnered from each set of partial contractions over batches $B_j$, as detailed in Sec. \ref{sec:background}. In contrast to this sequential approach, we introduce Non-Degenerate Batched Sampling (NBS), as illustrated in Fig. \ref{fig:algorithm_diagram} (right). In NBS, a user-defined number of bitstrings are collected from each batch $B_j$, such that degenerate work is not done resampling at these levels and as much of the space as the user desires can be explored.

We further develop two modes of such sampling: 1) non-proportional sampling, where we gather as much quantum data as possible (e.g., for downstream AI tasks), and 2) proportional sampling, where the original probability distribution of the circuit is preserved, maintaining the underlying quantum statistics.

For the proportional case, at each of the $f$ batches, $m_i$ shots are sampled from the conditional marginal distribution. For $B_1$, only one such contraction of $T^D$ via $P^1$ is required, at which point $m_i$ shots are sampled from the resulting probability vector. For $j>1$, however, all unique bitstrings $(s_1,\dots,s_{j-1})$ that have resulted from previous contractions must be sampled independently, as each of these partially projects the state into distinct conditional probability vectors. Such sampling branches to more or an equal number of distinct shot combinations with each batch, and thus we tend to incur an increasing number of distinct contractions with increasing $j$. Nevertheless, this is a strict improvement over the traditional trajectories methods, which always repeat all contraction steps, regardless of conditional bitstring overlap and the condition-free contraction of $B_1$.

For the non-proportional case, at each of the $f{-}1$ non-final batches, one or more shots are sampled from the conditional marginal distribution. Just like in the proportional case, these prefixes can branch and propagate forward, so the total shot count can grow combinatorially with increasing $j$ at a user-specified rate. However, non-proportional NBS also takes advantage of the fact that bitstrings harvested from the final batch $B_f$ are not used in subsequent shot calculations, meaning that we can gather many such shots from the final batch with little additional computational cost. We achieve this by offering multiple methods for extracting high volumes of training data from these final batches, including \textit{direct sampling} (probabilistic sampling from the $2^{b_f}$ population vector contracted from $B_f$) and \textit{exhaustive sampling}, where we gather all population entries above a user-specified threshold and returning them (optionally, with their respective probabilities). In this manuscript, all non-proportional PTSBE benchmarks are for exhaustive sampling.

In this manner, NBS allows us to gather massive amounts of noisy quantum data, both proportional, such that it adheres to traditional quantum statistics, or non-proportional, such that we produce maximal amounts of training data for downstream ML tasks.

\section{Experimental Setup}
\label{sec:setup}

\begin{figure}[t]
\centerline{\includegraphics[width=\columnwidth]{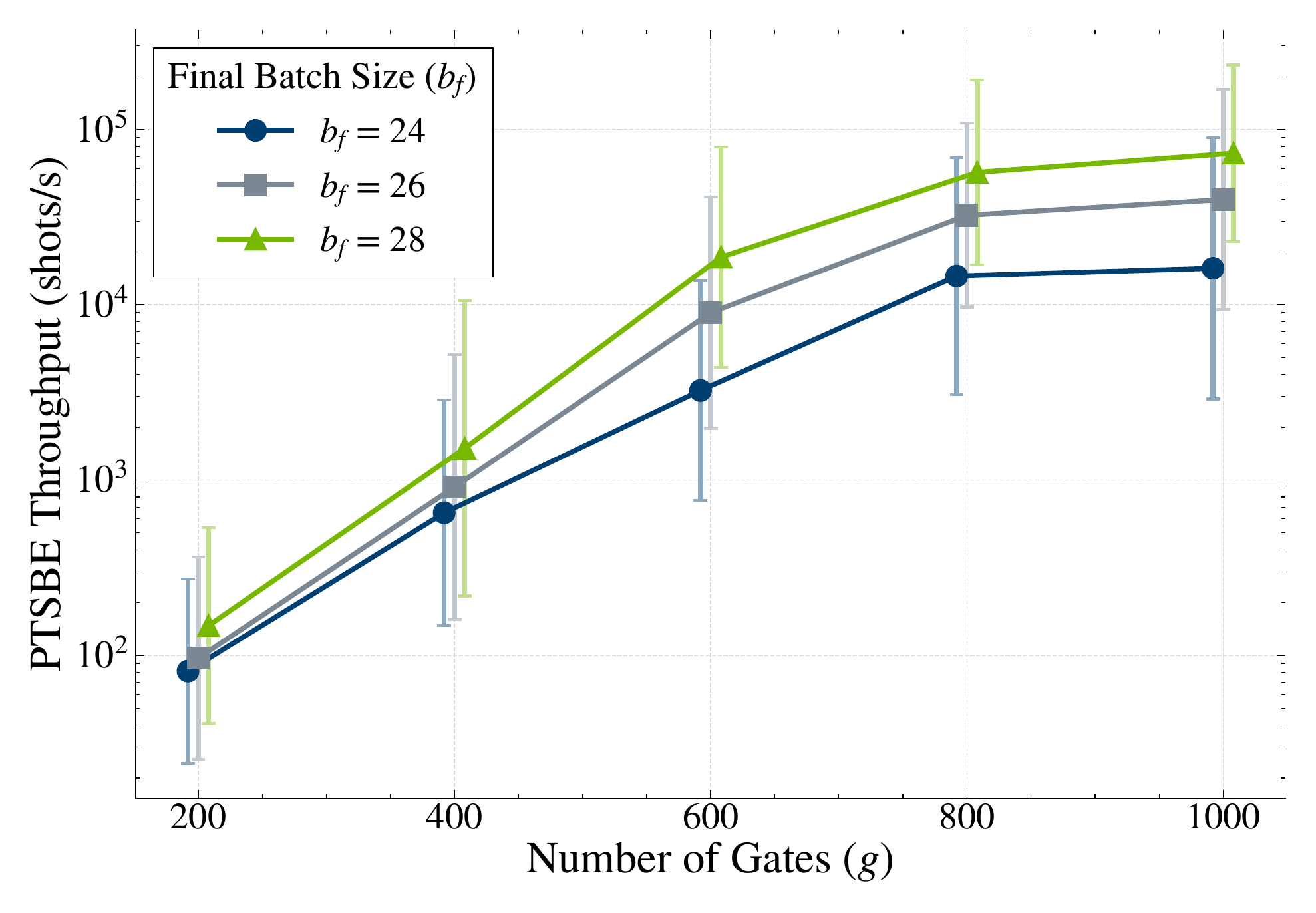}}
\caption{Data collection speedup vs gates vs final batch sizes $b_f$. Increasing the final batch size exponentially increases the size of the final Hilbert space ($\sim 2^{b_f}$), which is sampled directly and without need of subsequent contractions and thus contributes nearly proportionally to speedup. The extent of $b_f$'s influence on data collection rates is realized more fully for deeper circuits, due to the relationship between circuit depth and tensor rank.}
\label{fig:speedup_vs_final_batch_size}
\end{figure}

\subsection{Hardware and Software}

All experiments were conducted on NVIDIA H100 80GB GPUs (GH100 architecture) running CUDA v12.9.0 and with x86\_64 CPUs \cite{nvidia2022h100}. While this implementation was parallelized over distinct error sets $K_i$, we also detail the feasibility and benefits of an intra-error set multi-GPU implementation in Sec. \ref{sec:discussion}. The PTSBE implementation uses the cuQuantum (v26.01.0) cuTensorNet (v2.11.00) library \cite{bayraktar2023cuquantumsdkhighperformancelibrary} for tensor network contraction, with CuPy \cite{nishino2017cupy} (v2.2.3) as the GPU array backend and for CUDA event-based timing of contraction and path-finding phases. An intra-error set multi-GPU implementation would instead drop in cuPyNumeric \cite{cunumeric} rather than CuPy. Circuit gates were parsed into simulation input operands using Qiskit \cite{qiskit2024} as an intermediate representation. The CUDA-Q  (v0.13.0) baseline uses the CUDA-Q \texttt{tensornet} backend with the same hardware. Unless explicitly mentioned, default parameters for these packages were used. The numerical precision is set to \texttt{complex128}.

\subsection{Circuit Generation}

Random quantum circuits $T^D$ were generated with configurable qubit counts ($n \in \{50, 75, 100, 150, 200\}$) and gate counts ($g \in \{200, 400, 600, 800, 1000\}$). Each circuit consists of single-qubit gates ($H$, $X$, $Y$, $Z$, $T$, $R_x$) and two-qubit nearest-neighbor controlled gates ($CX$, $CY$, $CZ$, $CH$, $CR_x$), with $20\%$ of gates being two-qubit operations. As described in Sec. \ref{sec:methods} and shown in Fig. \ref{fig:contraction_path_reuse}, noise channels were coupled to each coherent gate. The noise gates were randomly drawn and included Pauli errors ($X$, $Y$, $Z$) for single-qubit gates and two-qubit depolarization \cite{nielsen2010quantum} for two-qubit gates, with error probabilities uniformly sampled on the interval $[0.02, 0.2]$.

\subsection{Experimental Parameters}
\label{subsec:experimental_parameters}

\begin{figure}[h!]
\centerline{\includegraphics[width=\columnwidth]{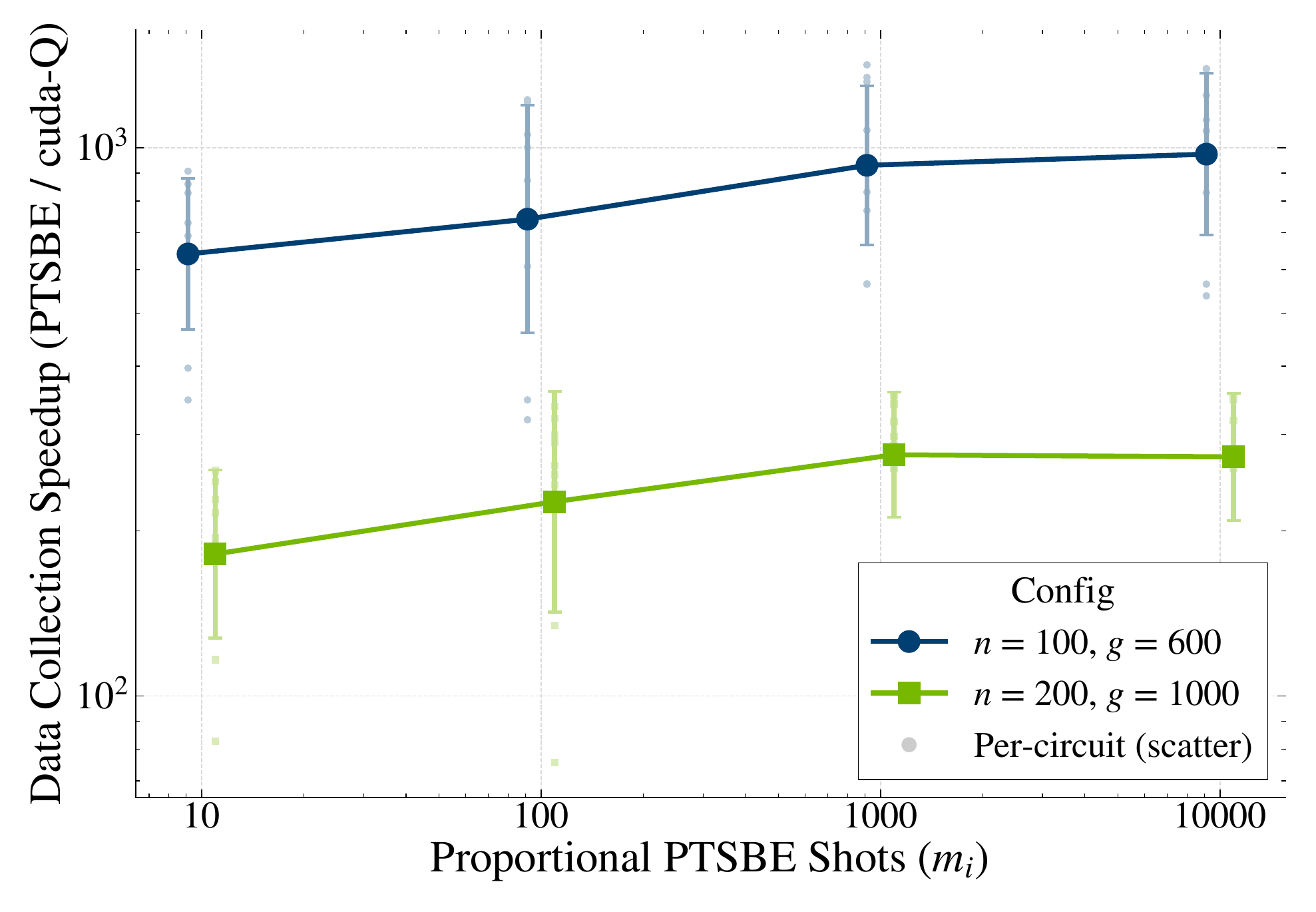}}
\caption{Proportional sampling data collection speedup for various circuit sizes and depths. As proportional sampling must adhere strictly to the statistics of a given circuit configuration, it cannot take advantage of large swaths of Hilbert space that non-proportional sampling can. Likewise, due to this strict sampling, the speedup is not particularly sensitive to shot number, as evidenced by the fact that speedup increases with larger $m_i$ remain within one standard deviation. However, it can still reach up to $\sim 1000\times$ speedup due to the advantages of UPV (see Sec. \ref{subsec:upv} and Fig. \ref{fig:path_vs_contraction}), as well as flexible and optimized batch sizes (see Fig. \ref{fig:contraction_time_vs_batch_size}).}
\label{fig:proportional}
\end{figure}

For each $n$-qubit, $g$-gate configuration, 10 random circuit instances were pre-generated. These same circuits were then reused across all experiments to ensure consistent comparisons, as distinct random circuit instances represent a the largest source of experiment variability. Unless otherwise noted, PTSBE experiments used a non-final batch size of $b = 10$ qubits with 1 shot per non-final batch, a final batch size of $b_f = 28$ qubits, and 100 hypersamples (contraction path optimization iterations). Traditional trajectories via CUDA-Q experiments have an inflexible batch size of $b=24$ and the number of hypersamples was set to 1, as this was found to be the optimal value through hyperparameter optimization.

\subsection{Metrics}

\begin{figure*}[h!]
\centerline{\includegraphics[width=\textwidth]{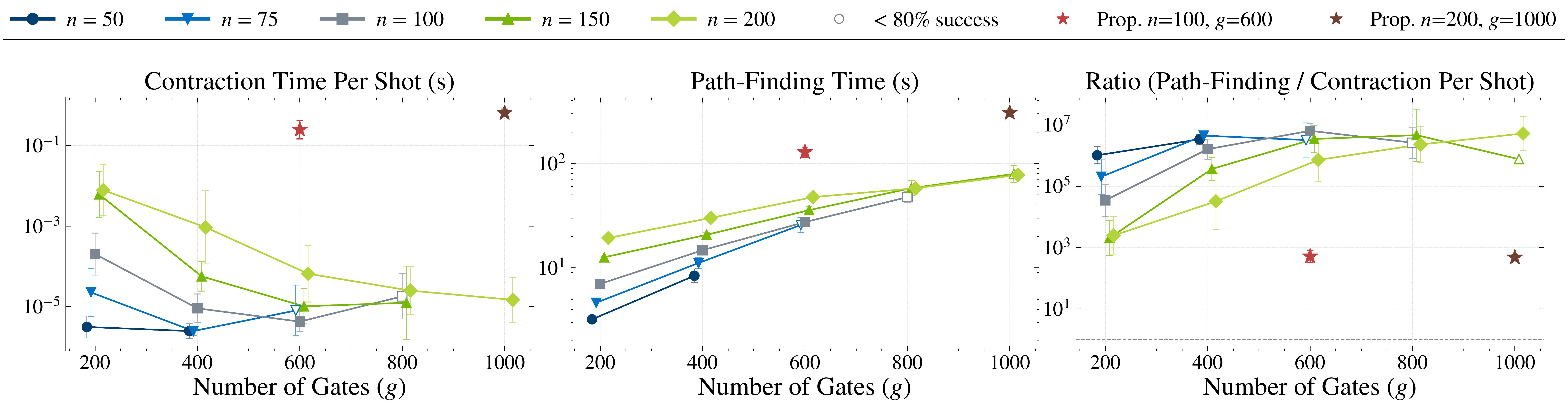}}
\caption{(Left) Contraction time per unique shot vs the number of gates $g$ for circuits with various numbers of qubits $n$. While tensor networks can handle circuits with large $n$, in many regimes their complexity grows exponentially with depth $g$. However, the populated states of such tensor networks also tend to grow, leading to lower per unique shot. (Center) The complexity of path finding grows with both $n$ and $g$. (Right) The ratio of path finding time to contraction time varied over the same set of parameters. The consistently high ratios indicate that repeated path finding (a subroutine whose cost is made negligible by our UPV technique) fundamentally limits the acceleration of unoptimized tensor network trajectory simulations. All connected markers are non-proportional PTSBE data, while the red and black star data points are from proportional PTSBE. In order to give the most conservative estimate of optimized PTSBE's capabilities possible, we use 1 hypersample per contraction path, as this increase contraction times and decreases path-finding times.}
\label{fig:path_vs_contraction}
\end{figure*}

We define \emph{throughput} as the total number of unique labeled bitstrings produced per unit of GPU utilization time (wall clock time) required to produce them. For optimized PTSBE, this is the total shot count across all $m_i$ error patterns produced from a single $K_i$ divided by the contraction loop time, which encompasses error injection, batch-wise contraction, and sampling for every error pattern. We note that contraction path finding is not included, as our UPV algorithm allows for the same path to be cached for arbitrarily many evaluations, rendering it negligible for HPC-level trajectory simulations, which by definition incorporate large numbers of error sets and very large shot volumes. For CUDA-Q trajectory simulations, throughput is the number of shots (quantum measurements) returned by \texttt{cudaq.sample()} per unit of execution time required to obtain them. As these CUDA-Q trajectory simulations lack any kind of path caching or batched sampling capabilities, they are universally slower, usually by orders of magnitude.

Finally, and most relevantly, we define the \emph{data collection speedup} as the ratio of PTSBE throughput to CUDA-Q throughput. For each benchmarking point in this work, we compute this speedup for each of 10 random circuit instances generated for each combination of qubit number $n$ and number of quantum gates $g$. All summary statistics, i.e., markers in figures and values reported in the text, are geometric means over these per-circuit throughput values, with error bars showing $\pm$1 geometric standard deviation. We use the geometric mean and standard deviation because throughput values span several orders of magnitude across configurations, and these metrics provide a meaningful central tendency on a logarithmic scale.

In data figures, solid markers indicate parameter configurations where $>80\%$ or random circuits have successfully completed and hollow markers indicate where $>20\%$ of random circuits have failed due to excessive contraction time or memory footprint. All such contraction failure ratios were the same for both implementations, indicating that the failure was universal to implementation type and thus inherent to the computational complexity of the circuit itself.

\section{Results}

Both the non-proportional and proportional implementations of optimized tensor network PTSBE demonstrate substantial speedups.

\subsection{Non-Proportional Sampling}

Due to non-proportional sampling's ability to extract all measurement outcomes of non-negligible probability from the final batch $B_j$, it demonstrates data collection speedups of up to $10^8$ times. As illustrated in Fig. \ref{fig:speedup_vs_gates}, the quantity of non-negligible measurement outcomes is a function of both the number of qubits $n$ and the number of gates $g$. This is because, when $g$ is large relative to $n$, we have a deep quantum circuit and a high-rank tensor network, wherein many such distinct quantum states of non-negligible probability exist and thus massive amounts of unique shots (unique quantum measurement data) are extracted.

As larger numbers of qubits $n$ take more gates to reach a deep circuit/high-rank state, we see a delay in data collection speedup to higher and higher $g$ for larger $n$. However, we also note that all $n$ explored converge to roughly the same data collection speedup $\sim 10^8$. This common data collection speedup is due to the choice of final batch size $b_f$, which roughly bounds the maximum number of extracted unique shots to be $\propto 2^{b_f}$. This effect of sample efficiency by final batch-size $b_f$ is demonstrated for $n=200$ systems in Fig. \ref{fig:speedup_vs_final_batch_size}, where $b_f=24$, $b_f = 26$, and $b_f = 28$ are separated by roughly a data collection speedup factor of $2{-}4$ times each. In this work, we cap $b_f$ at $28$ qubits as this is the largest size population vector that we can fit on a single H100 GPU, but future studies should explore larger values of $b_f$ with an intra-error set multi-GPU implementation as, presumably for large quantum systems, larger $b_f$ would yield even greater data collection speedups. Likewise, despite the suggestive relationship between data collection speedup and $b_f$ in Fig. \ref{fig:speedup_vs_final_batch_size}, we recognize the standard deviations between these two quantities are too large to serve as definitive proof. These large standard deviations are likely due to the large circuit-to-circuit complexity variations and shot-to-shot population differences. As a result, a more exhaustive study would be needed to definitively characterize the relationship between data collection speedup and $b_f$.

Finally, we highlight the large role that reusing contraction paths via UPV play in this speedup, as illustrated in Fig. \ref{fig:path_vs_contraction} (non-proportional sampling results are all the non-star markers). For all regimes explored, contraction time per second remains a fraction of a second, while path finding times are often tens of seconds, highlighting the importance of contraction path caching.

\subsection{Proportional Sampling}

We achieve proportional tensor network PTSBE sampling that is up to $1000 \times$ faster than the vanilla CUDA-Q implementation. Fig. \ref{fig:proportional} demonstrates that, while this speedup is highly dependent on circuit parameters (e.g., $n$ and $g$) it is largely independent of the number of shots $m_i$. Such minor dependence on $m_i$ stems from the lack of large and uniform sampling of the final batch $B_f$, which is not permitted in proportional sampling as it does not preserve the quantum statistics. Instead, all batches besides $B_1$ are carried out with some contraction overlap due to shared bitstrings $(s_1,\dots,s_{j-1})$, as detailed in Sec. \ref{subsec:nbs}. While we do note that this overlap (and thus the resulting speedup) would be considerably greater for non-random, structured quantum circuits, we acknowledge that contraction minimization with our NBS technique is of only minor utility in the random circuit case.

\begin{figure}[h!]
\centerline{\includegraphics[width=\columnwidth]{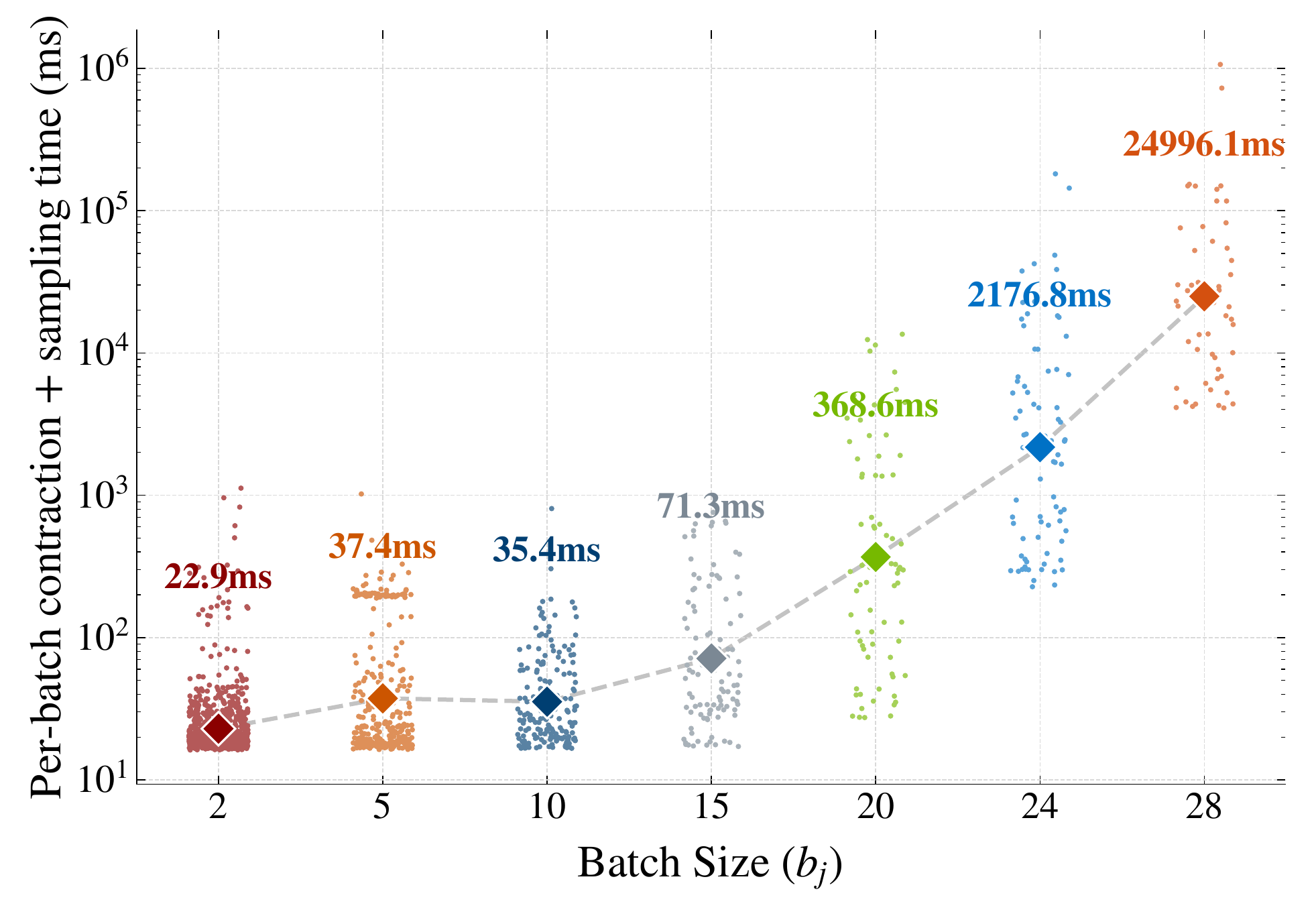}}
\caption{Contraction time per batch vs batch-size $b$ for $n=100$, $g=600$ circuits. Previous trajectories methods have fixed batch size (e.g., for CUDA-Q, $b=24$). While larger batch sizes do compute bitstrings for more qubits, requiring fewer contractions overall, such larger partial contractions are often much more costly. For example, at $b=24$, we require $\sim 90$ms per qubit contraction ($24$ qubits divided by $2176.8$ms), while we only require $\sim 3.5$ms per qubit contraction for $b=10$ ($10$ qubits divided by $35.4$ms). For this reason, smaller batch sizes are advisable for all non-final batches in high-throughput PTSBE and all batches in proportional PTSBE. This implementation flexibility was not exposed prior to our work, nor, to our knowledge, its impact on such simulations appreciated.}
\label{fig:contraction_time_vs_batch_size}
\end{figure}

Instead, the data collection speedup of our proportional sampling implementation stems from two main sources: 1) the elimination of path finding overheads via UPV and 2) the optimization of contraction parameters through the development and optimization of a more flexible trajectory methods interface. The utility of these two contributions can be understood through Fig. \ref{fig:path_vs_contraction} and Fig. \ref{fig:contraction_time_vs_batch_size}. As for the former, the red and black stars on Fig. \ref{fig:path_vs_contraction} correspond to $10{,}000$-shot proportional sampling for $n=100,g=600$ and $n=200,g=1{,}000$ circuits, respectively. In both cases, the contraction time per shot is less than a second while path finding time is in the hundreds of seconds. The substantial ratio of these two values alludes to the important role that UPV plays in optimized PTSBE speedup.

The remaining source of speedup can be understood through Fig. \ref{fig:contraction_time_vs_batch_size}. While traditional trajectory implementations, such as CUDA-Q, permit only a single, fixed batch size $b_j$, our implementation provides a flexible interface through which each $b_j$ can be set to user-specified values and thereby optimized. This allowed us to optimize over $b_j$ for the optimal rate of \textit{contracted qubits per unit time}. In Fig. \ref{fig:contraction_time_vs_batch_size}, we see that the default CUDA-Q value of $b_j = 24$ is actually an expensive batch-size selection, at least for the random circuits explored in this work. For $b_j = 24$, we glean a relatively slow rate of $24 / 2176.8 \text{ms} \approx 11$ contracted qubits per second. Conversely, $b_j=10$ is an approximate contraction-time minimum, obtaining a far more efficient rate of $10/35.4 \text{ms} \approx 282$ contracted qubits per second. We note that Fig. \ref{fig:contraction_time_vs_batch_size} also illustrates that in non-proportional sampling, where we have set $b_f = 28$, contracting over $B_f$ is by far the most expensive contraction. However, as shots from the final batch require no further data processing, the large number of unique shots gleaned from this step outweighs the cost of contraction.

The achievement of up to $1000 \times$ speedups for very \textit{general purpose} proportional sampling is a major boon for a wide array of quantum tensor network sampling procedures, not just those that incorporate non-proportional sampling. Even in these general regimes, our work provides a platform for the virtual elimination of contraction path finding's sizeable overhead and highlights the need for flexible and optimizable sampling interfaces.

\section{Discussion}
\label{sec:discussion}

In this manuscript, we have introduced three key innovations for tensor network simulations of noisy quantum systems: 1) an error-independent unified path variation scheme (UPV), 2) non-degenerate batched sampling (NBS), and 3) increased flexibility/studies in optimal tensor network sampling protocols. Not only do these innovations increase the data collection efficiency of non-proportional simulations by more than $10^8 \times$, they are also applicable for a wide variety of more traditional (proportional) sampling algorithms, accelerating these by up to $1000 \times$. These dramatic and widely applicable speedups represent a major boon to tensor network sampling for noisy quantum states, enabling ML-relevant data collection and even faithful statistical sampling for quantum states at previously intractable scales.

Although these findings are very promising, potential future work based on these innovations abounds. While the unified contraction path framework offered by UPV virtually eliminates contraction path overhead, its fixed application precludes the use of lightcone simplification, a useful tool for shallow and poorly connected tensor networks. Enabling lightcone simplification with UPV would require more advanced lightcone simplification algorithms that are able to track and structure error patterns. Likewise, despite the careful reutilization of both contraction paths (UPV) and partial shot information (NBS), a good deal of degenerate work is still preformed within the tensor contraction scheme itself. For instance, whole regions of partially contracted intermediate tensors may have no error changes between subsequent error sets $K_i$ and $K_{i+1}$, yet they are still prepared repeatedly. The PTSBE pre-trajectory error calculations would allows us to take advantage of these opportunities for intermediate caching, if our UPV framework were expanded to be contraction-intermediate aware. The execution order of pre-sampled error sets could even be optimized so as to maximize opportunities for intermediate caching.

\section*{Acknowledgment}

The authors acknowledge the following people for useful conversations: Dmitry Lyakh, Benedikt Kloss, Salvatore Mandra, Benjamin Villalonga, and Ali Charara.

\bibliographystyle{ACM-Reference-Format}
\bibliography{main}

\end{document}